\documentclass[aps,prl,showpacs,amsmath,amsfonts,amssymb,twocolumn]{revtex4} 
\usepackage{amsthm}
\usepackage{bbm}
\usepackage{psfrag}
\usepackage[dvips]{graphicx}
  
\newcommand\NN{{\mathbbm{N}}}

\newcommand\dd{{\mathrm{d}}}
\newcommand\ee{{\mathrm{e}}}

\DeclareMathOperator{\Heaviside}{\Theta}
\DeclareMathOperator{\Dirac}{\delta}

\newtheorem*{proposition}{Theorem}

\begin{document} 


\title{Phase transitions induced by saddle points of vanishing curvature} 

\author{Michael Kastner} 
\email{Michael.Kastner@uni-bayreuth.de} 
\affiliation{Physikalisches Institut, Universit\"at Bayreuth, 95440 Bayreuth, Germany} 

\author{Oliver Schnetz} 
\affiliation{\mbox{Institut f\"ur Theoretische Physik III, Friedrich-Alexander-Universit\"at Erlangen-N\"urnberg,}\\ Staudtstra{\ss}e 7, 91058 Erlangen, Germany} 

\date{\today}
 
\begin{abstract}
Based on the study of saddle points of the potential energy landscapes of generic classical many-particle systems, we present a necessary criterion for the occurrence of a thermodynamic phase transition. Remarkably, this criterion imposes conditions on microscopic properties, namely curvatures at the saddle points of the potential, and links them to the macroscopic phenomenon of a phase transition. We apply our result to two exactly solvable models, corroborating that the criterion derived is not only valid, but also sharp and useful: For both models studied, the criterion excludes the occurrence of a phase transition for all values of the potential energy but the transition energy. This result adds a geometrical ingredient to an established topological condition for the occurrence of a phase transition, thereby providing an answer to the long standing question of which topology changes in configuration space can induce a phase transition.
\end{abstract}

\pacs{05.70.Fh, 05.20.-y, 75.10.Hk} 

\maketitle 

A phase transition is an abrupt change of the macroscopic properties of a many-particle system under variation of a control parameter. Typical examples of phase transitions are the sudden disappearance of the electric resistance when cooling a superconducting material below its transition temperature, or the evaporation of a liquid at temperatures above its boiling point. Considerable progress has been made in the understanding of various aspects of phase transitions, in particular in the study of critical phenomena by renormalization group techniques. Other aspects are less understood. One of those is the question under which conditions on, say, the Hamiltonian of a system we can expect a phase transition to occur, and when can we exclude such a transition? Very little is known about conditions guaranteeing a phase transition to take place. The Peierls argument \cite{Peierls:36} or the Fr\"ohlich-Simon-Spencer bound \cite{FroeSiSpe:76} can be used to prove the existence of phase transitions without explicitly computing a thermodynamic potential, but the application of these methods is model specific and, depending on the system of interest, may be very difficult. There are a few remarkable results giving {\em necessary}\/ conditions for the occurrence of a phase transition, among those the Mermin-Wagner theorem \cite{MerWag:66} or the theorems on the absence of phase transitions in certain one-dimensional systems by \textcite{vanHove:50} and by \textcite{CuSa:04}. Such results are very useful since they can exclude the occurrence of phase transitions not only for specific models, but for {\em classes}\/ of systems.

One attempt to better understand the origin of phase transitions and the conditions under which they can occur dates back to \textcite{CaCaClePe:97}. These authors conjectured that phase transitions in classical many-particle systems were related to topology changes of the surfaces
\begin{equation}\label{eq:Sigma_v}
\Sigma_v=\left\{q\in\Gamma_N\,\big|\,V(q)=Nv\right\}
\end{equation}
of constant potential energy (per particle) $v=V/N$ in configuration space $\Gamma_N$, where $N$ is the number of degrees of freedom. In the ensuing years, this ``topological hypothesis'' was corroborated by a number of model calculations, although some peculiar (e.g., long-range) models disprove the all-encompassing validity of the conjecture (see \cite{Kastner:08} for a review). Only recently, a connection between phase transitions and topology changes in configuration space was demonstrated rigorously: For a certain class of systems with short-range interactions, a topology change of $\Sigma_v$ at $v=v_\text{t}$ was shown to be {\em necessary}\/ for a phase transition to take place at a transition potential energy $v_\text{t}$ \cite{FraPe:04,FraPeSpi:07}. This theorem, at least for the systems satisfying its assumptions, suggests the interpretation of certain topology changes as the relevant mechanism for the occurrence of a phase transition. Furthermore, in the manner of the other necessary conditions mentioned above, the theorem allows one to exclude a phase transition whenever such topology changes are absent.

Unfortunately, in many realistic situations an application of the theorem in \cite{FraPe:04,FraPeSpi:07} is impossible: For the typical systems of interest, the number of topology changes is very large, growing unboundedly with $N$, and in the thermodynamic limit $N\to\infty$ the topology changes become dense on a part or even the entire energy axis. So most of the topology changes turned out to be unrelated to phase transitions, but a criterion to identify the relevant ones was lacking. In this Letter, we supplement the topological approach to phase transitions with a geometrical ingredient which allows us to quantify the contribution of each topology change to thermodynamic quantities. Based on a finite-system result put forward in \cite{KaSchneSchrei:07}, we derive a topological-geometrical criterion by which we identify most of the topology changes as being unrelated to phase transitions in the thermodynamic limit. We apply this criterion to two exactly solvable models, the mean-field $XY$ model and the mean-field $k$-trigonometric model. For both models studied, our criterion excludes the occurrence of a phase transition for all but one value of the potential energy, coinciding precisely with the potential energy at which a phase transition occurs.

Under rather mild conditions on the many-particle potential $V$, topology changes of the subsets $\Sigma_v$ are closely related to saddle points of the $V$. This connection links the topological approach to phase transitions with the study of energy landscapes, an active field of research with many applications to clusters, bio-molecules, and glass-formers \cite{Wales}. We use the language of energy landscapes in the following exposition. 

{\em Preliminaries.}---We consider classical systems of $N$ degrees of freedom, characterized by a Hamiltonian function of standard form,
\begin{equation}\label{eq:Hamiltonian}
H(p,q)=\frac{1}{2}\sum_{i=1}^N p_i^2 + V(q), 
\end{equation}
where $p=(p_1,\dotsc,p_N)$ and $q=(q_1,\dotsc,q_N)$ are the vectors of momentum and position coordinates. The restriction to such a standard form is not essential, but simplifies the discussion: Since a quadratic form in the momenta $p_i$ yields only an irrelevant additive contribution to the free energy, it is sufficient to focus on the potential $V$ in the following. The potential $V$ is an analytic mapping from the continuous configuration space $\Gamma_N$ onto the reals, and in general $V$ has a number of {\em critical points}\/ (or {\em saddle points}) $q_\text{c}$ with vanishing differential, $\dd V(q_\text{c})=0$. We assume in the following that all critical points of $V$ have a non-singular Hessian ${\mathfrak H}_V$. In this case, $V$ is called a {\em Morse function}, and one can argue that the restriction to this class of functions is not a serious limitation: Morse functions form a dense, open subset of all analytic functions, and hence are {\em generic}. Therefore, even if $V$ is not a Morse function, it can be made into one by adding an arbitrarily small perturbation. A central proposition of Morse theory states that, for potentials $V$ being Morse functions, each topology change of the subsets
\begin{equation}
{\mathcal M}_v=\left\{q\in\Gamma_N\,\big|\,V(q)\leqslant Nv\right\}
\end{equation}
at some value $v=v_\text{c}$ corresponds to one or several critical points $q_\text{c}$ of $V$ with critical value $v_\text{c}=V(q_\text{c})/N$. The type of the topology change is determined by the {\em index}\/ of a critical point, i.e., the number of negative eigenvalues of the Hessian ${\mathfrak H}_V$ at $q_\text{c}$. As a consequence, the study of critical points and critical indices of $V$ is equivalent to the study of topology changes of ${\mathcal M}_v$. Since $\Sigma_v$ as defined in \eqref{eq:Sigma_v} is the boundary of ${\mathcal M}_v$, both subsets are closely related. We now continue the exposition in the language of saddle points and reinterpret our results in the language of topology changes only in the summary.

{\em Finite system nonanalyticities.}---In a recent Letter \cite{KaSchneSchrei:07}, we presented an exact, quantitative account of the contribution of a single critical point of a Morse function $V$ to statistical mechanical quantities of finite systems like the (configurational) density of states
\begin{equation}
\Omega_N(v)=\int_{\Gamma_N}\dd q \Dirac[V(q)-Nv]
\end{equation}
or, equivalently, the entropy $s_N(v)=\ln[\Omega_N(v)]/N$ per degree of freedom. We found that the density of states $\Omega_N=\Omega_N^\text{a}+\Omega_N^\text{na}$ can be written as the sum of a ``harmless'' analytic term $\Omega_N^\text{a}$ and a nonanalytic term which, to leading order, is given by
\begin{equation}\label{eq:Omega_na}
\Omega_N^\text{\rm na}(v)=\frac{(N\pi)^{N/2}}{N\Gamma(N/2)\sqrt{\left|\det\left[{\mathfrak H}_{V}(q_\text{c})/2\right]\right|}}h_{N,k}^\text{\rm na}(v)
\end{equation}
with the universal function
\begin{widetext}
\begin{equation}\label{eq:h_Nk}
h_{N,k}^\text{\rm na}(v)=\begin{cases}
(-1)^{k/2} \,v^{(N-2)/2} \Heaviside(v) & \text{for $k$ even,}\\
(-1)^{(k+1)/2} \,v^{(N-2)/2}\,\pi^{-1}\ln|v| & \text{for $N$ even, $k$ odd,}\\
(-1)^{(N-k)/2} (-v)^{(N-2)/2} \Heaviside(-v) & \text{for $N,k$ odd.}\\
\end{cases}
\end{equation}
\end{widetext}
Here $k$ denotes the index of the critical point and $\Heaviside$ is the Heaviside step function. This result, for its relation to the index of the critical point, is topological in nature, but it has also a geometrical ingredient: the determinant of the Hessian in the denominator of \eqref{eq:Omega_na} can be interpreted as some sort of curvature, quantifying the ``flatness'' of the critical point.

{\em Thermodynamic limit.}---We now take the above finite-system result as a starting point to compute the contribution of critical points of $V$ to thermodynamic quantities. To this purpose, we define the $\epsilon$-entropy in the thermodynamic limit as
\begin{align}\label{eq:sv}
s^{v_0,\epsilon}(v)&=\lim_{N\to\infty}\frac{1}{N}\ln\left[A^{v_0,\epsilon}_N(v)+B_N^{v_0,\epsilon}(v)\right]\\
&=\max\{a^{v_0,\epsilon}(v),b^{v_0,\epsilon}(v)\},\label{eq:sv_max}
\end{align}
where $B_N^{v_0,\epsilon}$ contains the nonanalytic contributions $\Omega_{N,q_\text{c}}^\text{na}$ from the critical points $q_\text{c}$ in the $\epsilon$-neighborhood of $v_0$,
\begin{equation}\label{eq:B_N}
B_N^{v_0,\epsilon}(v) = \sum_{\{v_\text{c}:|v_\text{c}-v_0|<\epsilon\}}\;\;\sum_{\{q_\text{c}:V(q_\text{c})=N v_\text{c}\}}\Omega_{N,q_\text{c}}^\text{na}(v),
\end{equation}
for some small $\epsilon>0$. The sum is written as a sum over critical values $v_\text{c}$ of $V$ and one over critical points $q_\text{c}(v_\text{c})$ of the respective critical value. Then a smooth function $A^{v_0,\epsilon}_N$ can be chosen such that the $\epsilon$-entropy $s^{v_0,\epsilon}$ coincides with the exact entropy $s=\lim_{N\to\infty}N^{-1}\ln\Omega_N$ in the interval $(v_0-\epsilon,v_0+\epsilon)$. From the results in \cite{FraPe:04,FraPeSpi:07} one can argue that, at least for well-behaved short-range systems, the analytic contributions in $A_N^{v_0,\epsilon}$ are irrelevant for phase transitions, converging uniformly to an analytic function $a^{v_0,\epsilon}=\lim_{N\to\infty}N^{-1}\ln A^{v_0,\epsilon}_N$ in the thermodynamic limit. Hence, we examine under which conditions
\begin{equation}
b^{v_0,\epsilon}(v)=\lim_{N\to\infty}\frac{1}{N}\ln B^{v_0,\epsilon}_N(v)
\end{equation}
may contribute to the entropy in the thermodynamic limit, thereby possibly inducing a phase transition.

Inserting the functional form of $\Omega^\text{na}_N$ into \eqref{eq:B_N}, we have to distinguish the various cases in Eq.\ \eqref{eq:h_Nk}. For $N=1\pmod 4$, we obtain
\begin{widetext}
\begin{equation}\label{eq:B_N_int}
B_N^{v_0,\epsilon}(v) = \frac{(N\pi)^{N/2}}{N\Gamma(N/2)} \left\{ \int_{v_0-\epsilon}^{v_0} \dd v'\left(v-v'\right)^{(N-2)/2} \left[{\mathcal N}_0(v')-{\mathcal N}_2(v')\right]
+ \int_{v_0}^{v_0+\epsilon} \dd v' \left(v'-v\right)^{(N-2)/2} \left[{\mathcal N}_1(v')-{\mathcal N}_3(v')\right]\right\},
\end{equation}
\end{widetext}
but other cases look similar and yield identical results in the thermodynamic limit. Here,
\begin{equation}\label{eq:N_j}
{\mathcal N}_\ell(v')=\sum_{q_\text{c}}J(q_\text{c}) \Dirac[V(q_\text{c})/N-v'] \Dirac_{\ell,k(q_\text{c})\;\text{mod 4}}
\end{equation}
are the distribution functions of the critical points $q_\text{c}$ with index $k(q_\text{c})=\ell\pmod 4$, weighted by their Jacobian determinant
\begin{equation}\label{eq:Jacobian}
J(q_\text{c})=\left| \det[{\mathfrak H}_V(q_\text{c})/2] \right|^{-1/2}.
\end{equation}
Under suitable assumptions on the thermodynamic limit properties of the relevant quantities we can evaluate the integrals in \eqref{eq:B_N_int} by Laplace's method and construct the upper bound
\begin{equation}\label{eq:b_bound}
b^{v_0,\epsilon}(v)\leqslant \frac{1}{2}\ln(\epsilon)+\sqrt{2\pi\ee}
+\max_{\ell,\, |v-v'|<\epsilon}\left[n_\ell+j_\ell(v')\right],
\end{equation}
where $\exp(N n_\ell)$ is the overall number of critical points with index $k=\ell\pmod 4$, generically growing exponentially with the number $N$ of degrees of freedom. The corresponding Jacobian density in the thermodynamic limit is given by
\begin{equation}\label{eq:j_ell}
j_\ell(v)
=\lim_{N\to\infty}\frac{1}{N}\ln\biggl(\sum\limits_{q_\text{c}\in Q_\ell(v,v+\epsilon)}\!J(q_\text{c})\;\bigg/ \!\!\sum\limits_{q_\text{c}\in Q_\ell(v,v+\epsilon)}1\biggr),
\end{equation}
where $Q_\ell(v,v+\epsilon)$ denotes the set of critical points $q_\text{c}$ with index $k(q_\text{c})=\ell\pmod4$ and with critical values $V(q_\text{c})/N$ in the interval $[v,v+\epsilon]$.

Now, if the $\max$-term in \eqref{eq:b_bound} is finite, we can always choose $\epsilon$ sufficiently small such that the bound for $b^{v_0,\epsilon}$ is smaller than $a^{v_0,\epsilon}$. Then, as a consequence of \eqref{eq:sv_max}, the entropy $s$ in the interval $(v_0-\epsilon,v_0+\epsilon)$ is exclusively determined by $a^{v_0,\epsilon}$, which we assume to be a smooth function. This argument can be repeated for any value of $v_0$, thereby demonstrating that the saddle point contributions in the $b$-term are harmless in the sense that they do not lead to nonanalyticities in the entropy. This reasoning, however, relies on the assumption that the $\max$-term in \eqref{eq:b_bound} is finite, and it breaks down in case $n_\ell$ or $j_\ell$ diverge. These observations lead us to the central result of this Letter.
\begin{proposition}\label{prop}
The saddle point contribution $b^{v_0,\epsilon}(v)$ cannot induce a phase transition at any potential energy in the interval $(v_0-\epsilon,v_0+\epsilon)$ if
\vspace{-1mm}
\begin{enumerate}
\item the number of critical points is bounded by\/ $\exp(CN)$ for some\/ $C>0$ and\label{cond1}
\vspace{-1mm}
\item the Jacobian densities have a thermodynamic limit of the form (\ref{eq:j_ell}) with\/ $j_\ell<\infty$ $\forall \ell\in\{0,1,2,3\}$\label{cond2}
\end{enumerate}
\vspace{-1mm}
inside the given interval.
\end{proposition}
In the literature on energy landscapes \cite{Wales}, the first condition is assumed to be true for generic many-particle systems, although an actual proof is missing. The second condition in the Theorem is the key result which allows us to identify most of the saddle points of the potential $V$, and therefore most of the topology changes in configuration space, as {\em irrelevant}\/ for the occurrence of a phase transitions. Only the ones which are sufficiently ``flat'' such that the Jacobian density $j_\ell$ diverges in the thermodynamic limit can give rise to a phase transition. To corroborate our reasoning and to illustrate the content of the Theorem, we present the computation of $j_\ell$ for two simple, exactly solvable spin models, showing that phase transitions occur precisely at the values of $v$ for which condition \ref{cond2} of the Theorem breaks down.

{\em Examples.}---The mean-field $XY$ model is a system of $N$ plane rotators described by angular variables $q_i\in[0,2\pi)$ ($i=1,\dotsc,N$). Each rotator is coupled to each other with equal strength $K>0$, where the interactions are described by the potential
\begin{equation}
V_{XY}(q)=\frac{K}{2N}\sum_{i,j=1}^N [1-\cos(q_i-q_j)]-h\sum_{i=1}^N \cos q_i.
\end{equation}
The rotators are subject to a magnetic field of strength $h$ which energetically favors orientations $q_i\approx \pi$ for $h<0$ and $q_i\approx 0 \pmod {2\pi}$ for $h>0$. In the limit $N\to\infty$ and $h\to0$, the system shows a continuous phase transition.

\begin{figure}[b]
\psfrag{v}{$\scriptstyle v$}
\psfrag{j_l}{$\scriptstyle j_\ell$}
\psfrag{j_ell}{}
\psfrag{v}{$\scriptstyle v$}
\psfrag{j_l}{$\scriptstyle j_\ell$}
\includegraphics[height=4.45cm,angle=270]{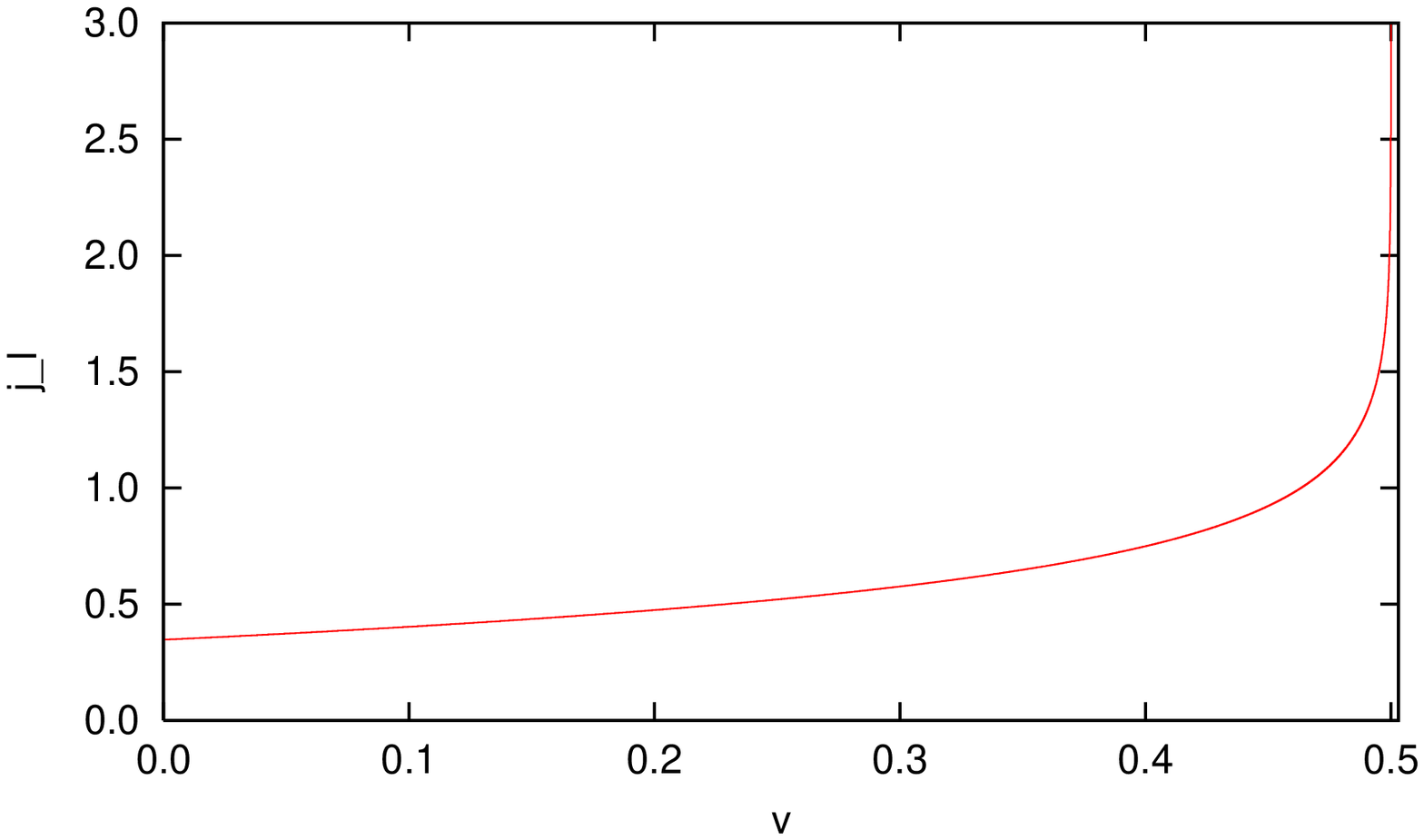}
\hspace{-5mm}
\includegraphics[height=4.45cm,angle=270]{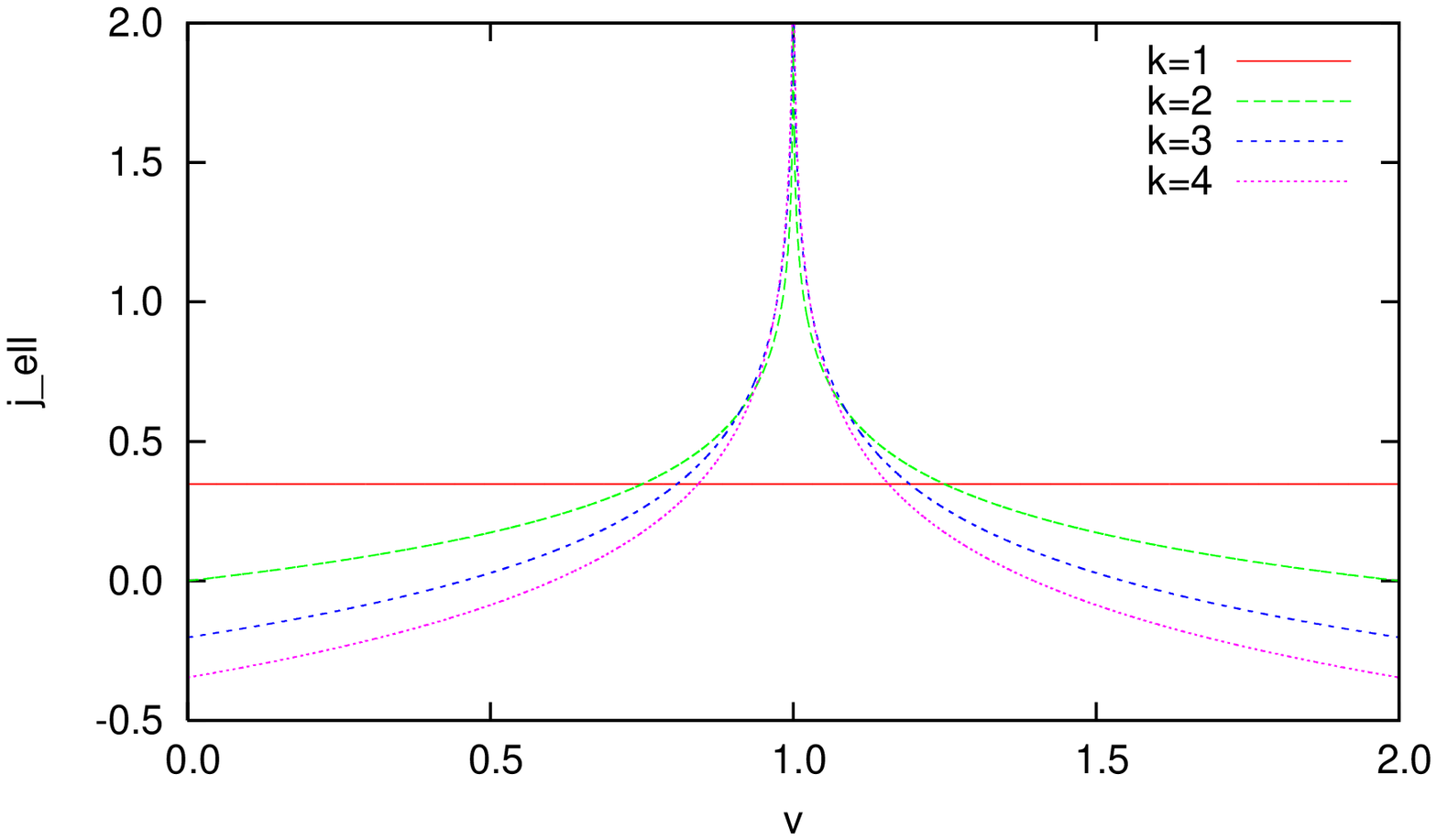}
\caption{\label{fig:j_ell}
(Color online) Plot of the graphs of the Jacobian densities $j_\ell$ as functions of the potential energy $v$ for the mean-field $XY$ model with coupling $K=1$ (left) and for the mean-field $k$-trig\-o\-no\-met\-ric model with $k=1,2,3,4$ and $\Delta=1$ (right). In agreement with our Theorem, the transition potential energies $v_\text{t}=1/2$ and $v_\text{t}=\Delta$ of these models coincide with the singularities of $j_\ell$.
}
\end{figure}

The critical points of this model and their indices have been analyzed by \textcite{CaPeCo:03} for arbitrary numbers $N$ of rotators. Critical points of $V_{XY}(q)$ were found to occur for $q=q_\text{c}\in\{0,\pi\}^N$, i.\,e., for $(q_1,\dotsc,q_N)$ with all components $q_i$ being either $0$ or $\pi$. Hence, in accordance with condition \ref{cond1} of the Theorem, the number of critical points is growing exponentially with $N$. The elements ${\mathfrak H}_{ij}=\partial^2 V/(\partial q_i \partial q_j)$ of the Hessian are also reported in \cite{CaPeCo:03}. Evaluated at a critical point, the diagonal elements of the Hessian can be written as
\begin{equation}
{\mathfrak H}_{ii}(q_\text{c})= \left[K\left(1-\frac{2n_\pi}{N}\right)+h\right]\cos q_i-\frac{K}{N},
\end{equation}
where $n_\pi$ is the number of $\pi$'s in the sequence $(q_1,\dotsc,q_N)\in\{0,\pi\}^N$. It has been shown in Appendix A.1 of \cite{CaPeCo:03} that, in the thermodynamic limit $N\to\infty$, the contribution of the off-diagonal elements to the eigenvalues of ${\mathfrak H}$ vanishes. Therefore, for large $N$ we can write the Jacobian determinant \eqref{eq:Jacobian} at a critical point $q_{\text{c}}$ as
\begin{equation}
J(q_{\text{c}})\!\xrightarrow{N\gg1}\! 2^{N/2}\biggl| \prod_{i=1}^N {\mathfrak H}_{ii}(q_{\text{c}}) \biggr|^{-1/2}\!\!\!\!=\left| \frac{K}{2}-\frac{Kn_\pi}{N}+\frac{h}{2} \right|^{-N/2}\!\!.
\end{equation}
Inserting an expression of $n_\pi$ depending on the potential energy per degree of freedom $v$,
\begin{equation}
\frac{n_\pi(v)}{N}\xrightarrow{N\gg1}\frac{1}{2K}\left(K+h\pm\sqrt{K^2+h^2-2Kv}\right),
\end{equation}
we can calculate the Jacobian density \eqref{eq:j_ell} for $N\to\infty$,
\begin{equation}
j_\ell(v)=\frac{1}{2}\ln 2-\frac{1}{4}\ln[K(K-2v)],\qquad\ell=0,1,2,3,
\end{equation}
in the limit $h\to0$. The graph of this function is plotted in Fig.~\ref{fig:j_ell} (left).
As is easily recognized, $j_\ell(v)$ is singular only at $v=K/2$, which is precisely the potential energy per degree of freedom at which a phase transition occurs for the mean-field $XY$ model with zero external field. In agreement with our Theorem, no phase transition is induced by the critical points with finite Jacobian density.

The mean-field $k$-trigonometric model, characterized by the Hamiltonian function
\begin{equation}
V_k(q)= \Delta N^{1-k}\sum_{i_1,\dotsc,i_k=1}^N [1-\cos(q_{i_1}+\dotsb+q_{i_k})],
\end{equation}
is another spin models consisting of plane rotators, with $\Delta$ determining the coupling strength between the rotators. For this model, the thermodynamic behavior and the critical points of the potential have been studied by \textcite{Angelani_etal:05}. The parameter $k\in\NN$ in the Hamiltonian function crucially determines the thermodynamic behavior of the model: In the limit $N\to\infty$, the system shows a discontinuous phase transition for $k\geqslant3$, a continuous one for $k=2$, and no phase transition for $k=1$. The results on the critical points of $V_k$ in \cite{Angelani_etal:05} can be assembled, in a calculation very similar to the one for the mean-field $XY$ model, to obtain the Jacobian densities $j_\ell$ as functions of the potential energy per particle $v$,
\begin{equation}
j_\ell(v)=\frac{1}{2}\ln\left(\frac{2}{\Delta k}\left|1-\frac{v}{\Delta}\right|^{1/k-1}\right),\quad\ell=0,1,2,3.
\end{equation}
The graph of this function is plotted in Fig.~\ref{fig:j_ell} (right) for $k=1,2,3,4$. $j_\ell$ is constant (and therefore bounded above) for $k=1$, in agreement with the absence of a phase transition in this case. For $k\geqslant2$ the Jacobian density $j_\ell(v)$ shows a divergence at $v=\Delta$, coinciding precisely with the transition potential energy of the phase transition of the mean-field $k$-trigonometric model and illustrating nicely the content of the Theorem. In contrast to short-range systems, phase transitions in long-range systems {\em need not be}, but {\em can be}\/ induced by saddle points of $V$ with diverging $j_\ell$. According to our results, this is the relevant mechanism for the two models studied.

{\em Summary.}---Based on a result describing the effect of saddle points of the potential $V$ on the finite-system entropy, we derived a necessary criterion for the occurrence of a phase transition in the thermodynamic limit. This criterion imposes conditions on microscopic properties, namely curvatures at the saddle points, and links them to the macroscopic phenomenon of a phase transition. Interpreting the result (via Morse theory) in terms of topology changes in configuration space, an answer is provided to the long standing question of which topology changes of $\Sigma_v$ or ${\mathcal M}_v$ can induce a phase transition. Applications to the mean-field $XY$ and $k$-trigonometric models show that the criterion is not only valid, but also sharp and useful: For both models, the criterion excludes the occurrence of a phase transition for all values of the potential energy but the transition energy.

Our Theorem provides a criterion to single out those saddles of the potential energy which are relevant for phase transitions, and it facilitates the study of phase transitions from the alternative point of view of energy landscapes. This approach is of particular interest for applications to the folding transition of proteins: energy landscape methods are frequently used for studying dynamical properties of such transitions, and our results allow investigation of both, dynamical and thermodynamical properties within a unified framework based on energy landscapes.

\bibliography{JacobianPT.bib}

\end{document}